\documentclass{mem}
\usepackage{amsmath}
\usepackage{natbib}
\usepackage{txfonts}
\usepackage{balance}
\usepackage{graphicx}
\usepackage{flushend}
\usepackage[breaklinks,pdftex]{hyperref}
\idline{75}{282}
\newcommand{\cms}{~cm~s$^{-1}$}
\newcommand{\kms}{~km~s$^{-1}$}
\newcommand{\nqso}{N_{\rm QSO}}
\newcommand{\zqso}{z_{\rm QSO}}
\newcommand{\lcdm}{$\Lambda$CDM}

\begin{document}

\title{
Spectrographs and Spectroscopists for the Sandage Test
}
   \subtitle{}
\author{
S. \, Cristiani\inst{1},
K. \, Boutsia \inst{2},
G. \, Calderone \inst{1},
G. \, Cupani \inst{1},
V. \, D'Odorico \inst{1},
F. \, Fontanot \inst{1},
A. \, Grazian \inst{3},
F. \, Guarneri \inst{1,4},
C. \, Martins \inst{5},
L. \, Pasquini \inst{4},
M. \, Porru \inst{1},
E. \, Vanzella \inst{6}.
          }
\institute{
Istituto Nazionale di Astrofisica --
Osservatorio Astronomico di Trieste, Via Tiepolo 11,
34131 Trieste, Italy,  
\email{stefano.cristiani@inaf.it}
\and
Las Campanas Observatory, Carnegie Observatories, 
Colina El Pino, Casilla 601, La Serena, Chile
\and
Istituto Nazionale di Astrofisica --
Osservatorio Astronomico di Padova, Vicolo dell'Osservatorio 5,  35122 Padova, Italy
\and
ESO - European Southern Observatory, Karl-Schwarzchild-Strasse 2, 85748 Garching bei München, Germany
\and
CAUP and IA-Porto, Rua das Estrelas s/n, 4150-762 Porto, Portugal
\and
 INAF – OAS, Osservatorio di Astrofisica e Scienza dello Spazio di Bologna, Via Gobetti 93/3, 40129 Bologna, Italy
}

\authorrunning{Cristiani }

\titlerunning{The Sandage test of the Redshift Drift}

\date{Received: Day Month Year; Accepted: Day Month Year}

\abstract{
The redshift drift is a small, dynamic change in the redshift of objects following the Hubble flow. Its measurement provides a direct, real-time, model-independent mapping of the expansion rate of the Universe.
It is fundamentally different from other cosmological probes: instead of mapping our (present-day) past light-cone, it directly compares different past light-cones. Being independent of any assumptions on gravity, geometry or clustering, it directly tests the pillars of the \lcdm\ paradigm. Recent theoretical studies have uncovered unique synergies with other cosmological probes, including the characterization of the physical properties of dark energy.
At the time of the original proposal by Sandage (1962) the expected change in the redshift of objects at cosmological distances appeared to be exceedingly small for reasonable observing times and beyond technological capabilities.
In the last decades progress in the spectrographs (e.g. ESPRESSO), in the collecting area of telescopes and in the samples of cosmic beacons, enabled by new datasets and new machine-learning-based selections, have drastically changed the situation, bringing the Redshift Drift Grail within reach. As a consequence, this measurement is a flagship objective of the Extremely Large Telescope (ELT), specifically of its high-resolution spectrograph, ANDES.
}
\maketitle{}

\section{Introduction}
Let me begin this contribution with a little tribute to Margherita Hack, 
recalling an episode of my adolescence.
My interest in Astronomy had just awakened,
but in the 70s and in a provincial town 
it was not so easy to find material to study it. 
In practice there was just a magazine, Coelum\footnote{http://www.coelum.com/} and  
few popular books, among which I recall "Modern Cosmology" \citep{Sciama71} and "Beyond the Moon" \citep{Maffei73}. So I wrote various letters to famed Italian astronomers asking for help, 
and Margherita was kind enough to send me her Astronomy course \citep{Hack73}, 
which I eagerly read.
Therefore she has some responsibility for instilling in me love for Astrophysics
and in particular for spectroscopy.
She would not miss the opportunity to emphasize that {\it "the study of the spectra of celestial bodies has been of fundamental importance for the birth of astrophysics. It is only thanks to the analysis of the radiations emitted by celestial bodies that it was possible to determine their surface temperature and density, their chemical composition, the motions of gases in stellar atmospheres and of stars in galaxies..."}\citep[Ch.2]{Hack1998}, which was not a trivial statement when she started her career. In fact, {\it "there had been an eclipse of Astrophysics and the eleven Italian observatories (Turin, Milan, Padua, Trieste, Bologna, Florence, Rome, Teramo, Naples, Catania and Palermo), plus the Carloforte station, were practically all directed by mathematicians who, in addition to doing research that was obsolete, were also very autocratic…
The positions of the stars, the constellations, the measurements of the parallaxes were studied. There was nothing physical." }\footnote{L. Bonolis, "Colloquio con Margherita Hack", 8 aprile 2003}
The enthusiasm for spectroscopy was reiterated during my thesis at the Asiago observatory by Augusto Mammano, one of the discoverers of the signature of the first microquasar in SS433 \citep{Mammano1980} as well as an avid  competitor in Margherita’s volleyball matches\citep[Ch.3]{Hack1998}.

\section{Optical Spectrographs (an ESO-biased view)}
So be it, Spectroscopy:
from the Boller \& Chivens Spectrograph at the Asiago Observatory
\citep{Barbieri1977, Barbieri1980}, to Boller \& Chivens Spectrographs at the La Silla Observatory \citep{Zeilinger1991}. In La Silla I had the privilege to witness and 
participate in the “fast and furious” growth \citep{Sandro2018} of new instruments, in particular spectrographs, 
in the two decades 1978-1998: the Coudé Échelle Spectrograph \citep[CES]{Enard1978}
in the Coudé room of the 3.6m telescope, initially fed by the 1.4m Coudé Auxiliary Telescope (CAT) and, a few years later, by an optical fibre from the 3.6m, CASPEC \citep{Sandro1983}, an efficient crossdispersed echelle spectrograph, and the ESO Faint Object Spectrograph \& Camera \citep[EFOSC]{Enard1984, Sandro1986} and the ESO Multi-Mode Instrument \citep[EMMI]{Dekker1986}
at the New Technology Telescope.
The high efficiency of the optics, coupled with the high quantum sensitivity of the newly introduced CCDs, boosted the performance of these instruments and, for the first
time, gave European astronomers the possibility of competing on crucial observing
modes with their colleagues at other 4m-class telescopes worldwide \citep{Sandro2018}.
Multiple-object spectroscopy (MOS) was inaugurated on relatively small fields with EFOSC thanks to the masks of the PUMA punchmachine\citep{Sandro1987}
and on larger fields taking advantage of optical fibers 
with the Fibre Optics Multi-Object Spectrograph \citep[OPTOPUS]{Enard1983, Cristiani1987}, both at the Cassegrain focus of the 3.6m telescope.

The list is non-exhaustive and limited to 1998; later on more (high-resolution) spectrographs would arrive in La Silla, such as FEROS \citep{FEROS1998} 
and HARPS \citep{HARPS2003} (and in the future SoXS \cite{SOXS2022} will come).
More details on these and many more ESO instruments can be found in \cite{Madsen2012}.

In 1999 the VLT era began and a new panoply of powerful spectrographs gradually became available,
making Paranal the most productive ground-based observatory:
FORS \citep{FORS1998}, 
UVES \citep{UVES2000}, 
FLAMES \citep{Pasquini2002},
VIMOS \citep{VIMOS2003}, 
X-shooter \citep{XShooter2011}, 
MUSE \citep{MUSE2010}, 
ESPRESSO \citep{ESPRESSO2021}.

An innovative pattern for the construction of ESO instruments was devised,
with the majority of them built by consortia of institutes,
 within a set of
standardized specifications. Institutes were rewarded for the costs incurred by them both in terms of staff and sometimes also in terms of hardware with 
observing nights (called Guaranteed Time Observations, GTO).

These collaborations provided advantages for both ESO and the national institutes, 
enabling an ambitious instrumentation programme, giving access to unique expertise nurtured in national
institutes and fostering a sense of ownership
of the VLT program in a significant fraction of the astronomical community.
For the institutes it led to the creation of competent, multidisciplinary
instrument teams around an ambitious project, and made it easier to obtain
funding from national agencies to develop the necessary infrastructure,
including integration and testing facilities \citep{Sandro2018}.
\begin{figure*}
\resizebox*{\hsize}{10cm}{\includegraphics[clip=true]{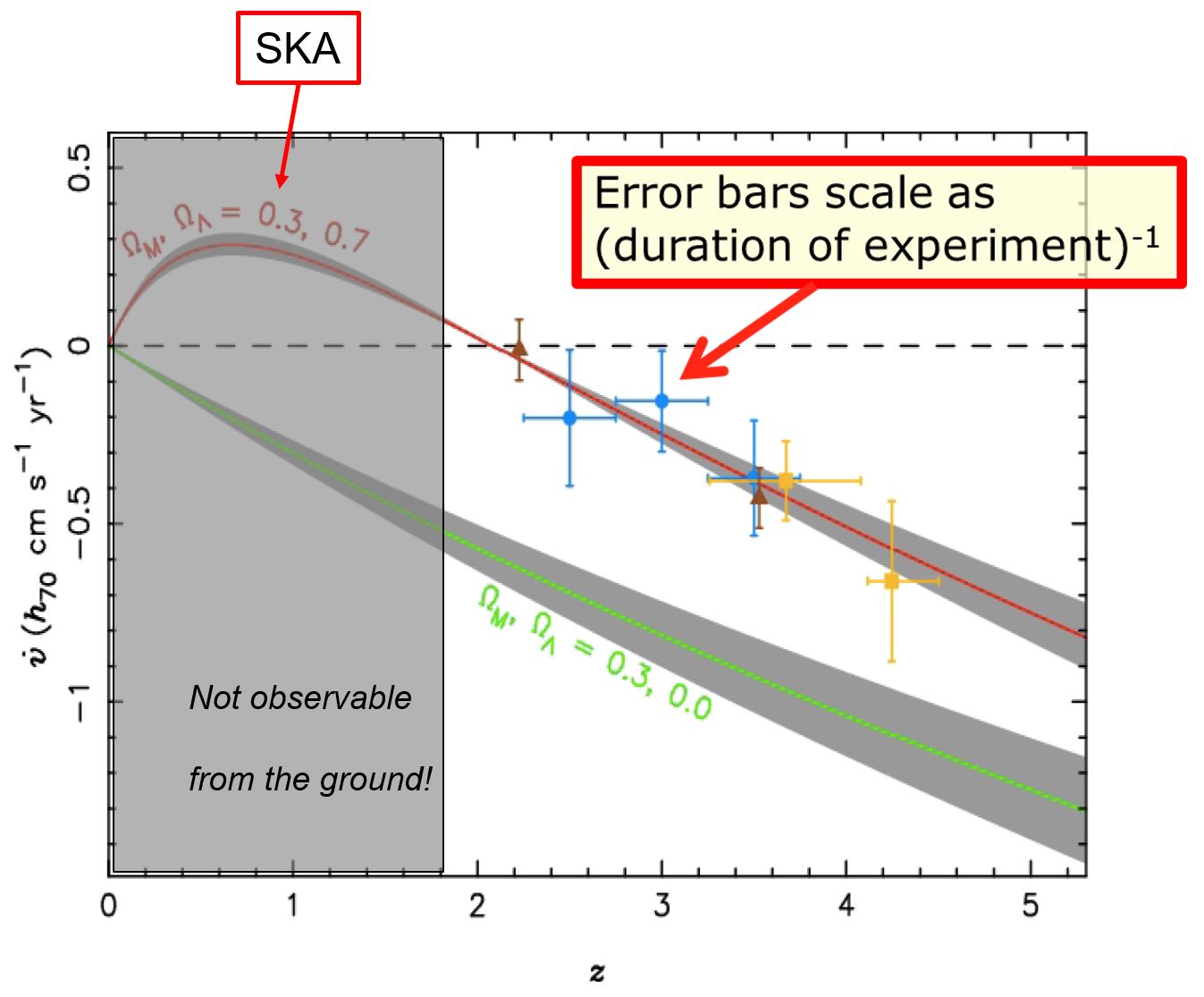}}
\caption{\footnotesize
Simulations of the Sandage test of the redshift drift 
using different example implementations of the experiment
over an interval of time of 20 yr with a 42-m telescope.
Blue dots: 20 quasars, binned into three redshift bins, equal time allocation, that provide the most precise measurement of $\dot z$. Yellow squares: selection of two higher redshift bins, 10 quasars maximizing the significance of the detection of a non-zero drift. Brown triangles: 2 quasars at lower redshift that provide the best combined constraint on $\Omega_{\Lambda}$. The grey shaded areas result from varying Ho by $\pm 8$ \kms Mpc$^{-1}$
(Adapted from \cite{Liske2008}).
}
\label{fig:SimSandage}
\end{figure*}
The momentum gained in this way extends to the future, as shown by the presentations at this conference about CUBES at the VLT (Covino et al.) and ANDES at the ELT (Marconi et al.).
\begin{figure}
\resizebox{\hsize}{!}{\includegraphics[clip=true]{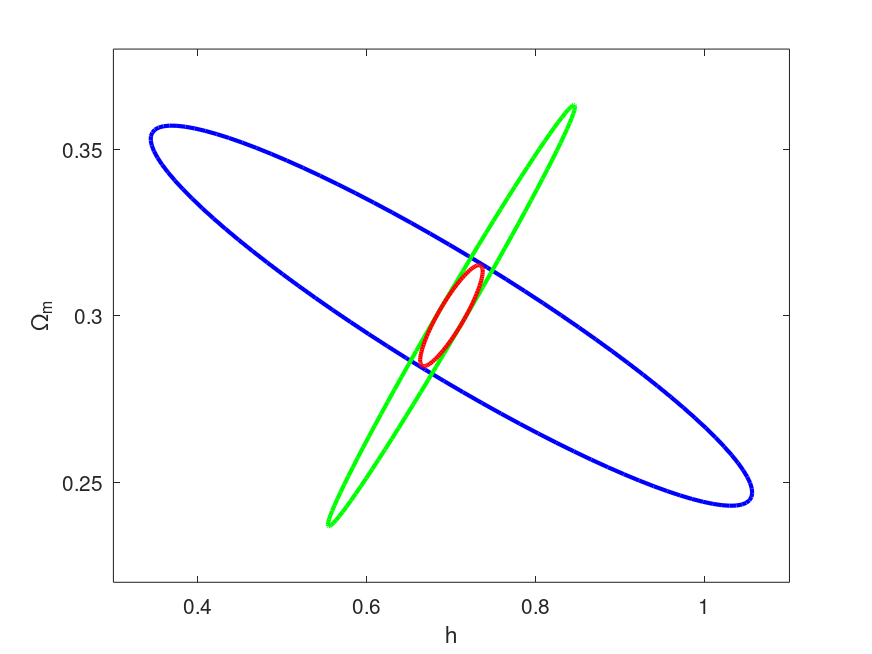}}
\caption{\footnotesize
Illustrating the synergy of redshift drift measurements by ELT/ANDES
(blue) and the SKA Observatory (green): the plot shows the constraints from redshift drift measurements alone: a flat LambdaCDM fiducial model was assumed,
but there are no external priors. The red contour is the combined constraint. 
}
\label{fig:DriftSynergy}
\end{figure}
\section{The Sandage Test of the Redshift Drift (a personally biased view)}

Around 2000 at ESO Garching two post-docs, Andrea Grazian and Eros Vanzella, triggered by a conversation with Luca Pasquini, approached me,
asking about the possibility of using UVES spectra to measure the variation of the expansion rate of the Universe. I told them that it was an interesting idea, that more than a decade before Peter Shaver and I had fancied about applying radio observations to measure the effect, that it was instructive to work out the math, but the final result would be that the expected signal was beyond the possibility of detection with the technology of the time.

Andrea and Eros did their homework,
\begin{equation}
\frac{dz}{dt_o} = (1+z) ~H_o - H(z)
\end{equation}
and stubbornly continued to think about ways to perform this measurement, for example taking advantage of the delay time caused by gravitational lensing. We found that in 1998 Loeb had written a fundamental paper \citep{Loeb1998} proposing to use the Lyman Forest observed in quasar spectra for this measurement and also that the original idea dated back to 1962 with a seminal paper by \cite{Sandage1962}.
\begin{figure*}
\resizebox{\hsize}{!}{\includegraphics[clip=true]{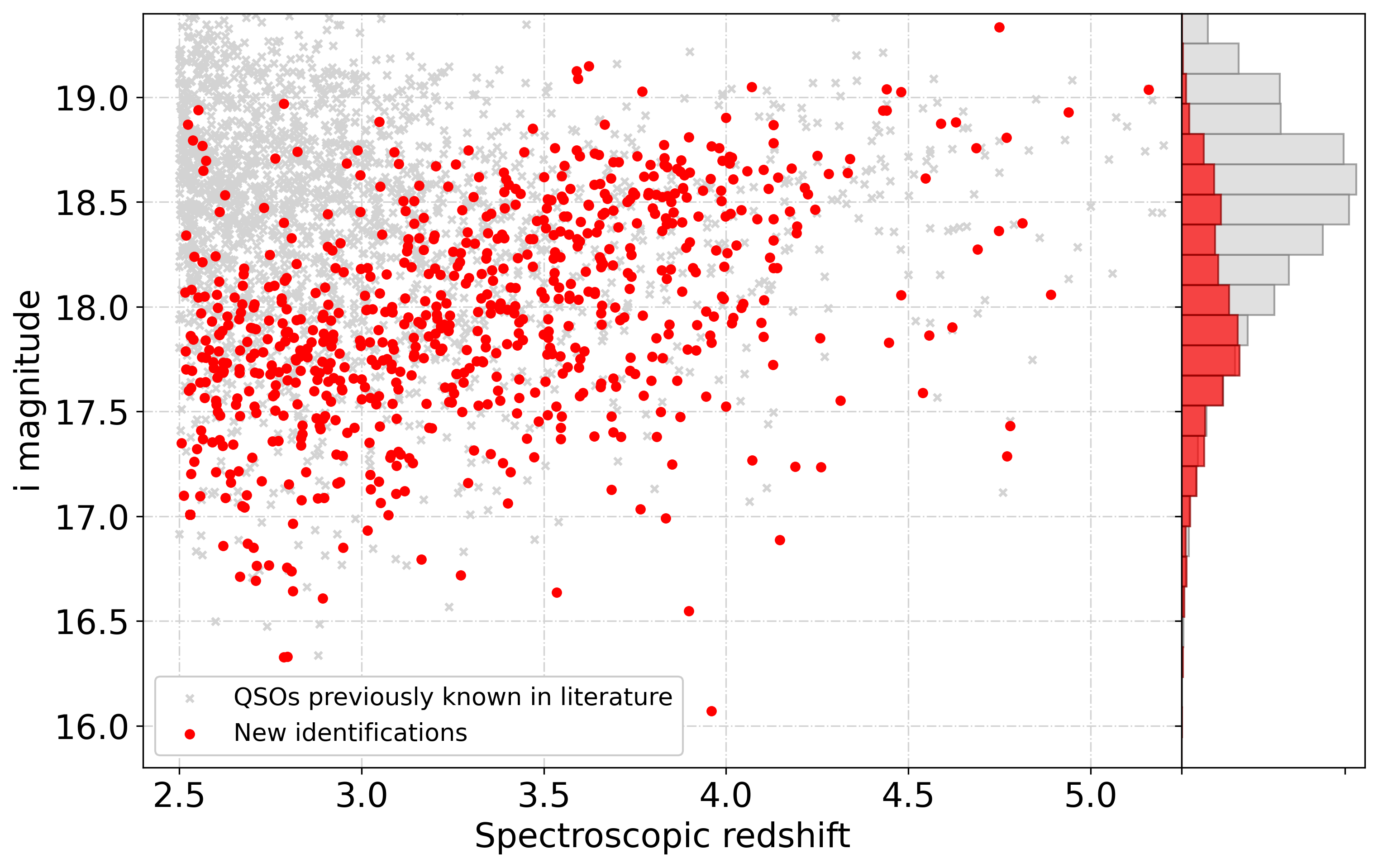}}
\caption{\footnotesize
i-band magnitudes versus spectroscopic redshifts
for $z > 2.5$ quasars in the Southern hemisphere. Quasars
discovered by QUBRICS have been highlighted in red.
}
\label{fig:z_mag_QSO}
\end{figure*}
It was only few months later, under a shower and with my subconscious nourished by the visionary proposal of OWL \citep{OWL2000}, that I suddenly realized: "I told Andrea \& Eros the detection is not possible with the {\bf present} technology, but what about {\bf future} technology?
Particle physicists are building the LHC, we may have OWL".
In this way we got into the Redshift Drift business,
joining other enthusiasts of this experiment.
To carry it out a new spectrograph, CODEX, was envisaged for OWL \citep{Pasquini2005} and indeed the redshift drift became one of the four key scientific cases of the new European Extremely Large Telescope (E-ELT).
The CODEX experiment is conceptually
very simple: by making observations
of high redshift objects with a time interval
of several years, we want to detect
and use the wavelength shifts of spectral
features of light emitted at high redshift
to probe the evolution of the expansion of
the Universe directly \cite{Cristiani2007,Liske2008}.
Being independent of any assumptions on gravity, geometry or
clustering, the redshift drift directly tests the pillars of the \lcdm paradigm. Recent theoretical studies have uncovered unique synergies with other cosmological probes, including the characterization of the physical properties of dark energy \cite{Martins2016, Esteves2021}.
The signal has units of acceleration and is expected to be extremely small (ca. 5 \cms per decade) but grows
linearly with time. Its detection requires several epochs of observations, extremely stable wavelength calibration
($ \Delta\lambda / \lambda \sim 10^{-10}$ or, equivalently, 3 \cms per decade), and high signal-to-noise (SNR) observations. 
Accuracies not far from what we need for detecting the cosmic signal are  being reached in the observations of radial velocity
perturbations induced by extra-solar planets in stellar spectra (e.g. ESPRESSO~\cite{ESPRESSO2021}).
We want to do the same but with objects that are hundred thousand times fainter than the extra-solar planets targets, and on timescales of decades.
An extremely large light bucket is needed and 
an absolute calibration source accurate on long time scales, the Laser Frequency Comb 
\citep[LFC]{Murphy2007}.

The \cite{Liske2008} paper concluded that
a 42-m telescope is capable of unambiguously
detecting the redshift drift over a period of $\sim 20$ yr using 4000 h of observing time and, on the basis of detailed simulations, that a precision of
\begin{flalign}
\sigma_v & = 1.35 \mbox{\cms} ~ \times \\ \nonumber
& \times \left(\frac{\mbox{S/N}}{2370}\right)^{-1} \!
\left(\frac{\nqso}{30}\right)^{-\frac{1}{2}} \!
\left(\frac{1 + \zqso}{5}\right)^{-1.7} 
\end{flalign}
can be achieved, depending on the redshift and the number of quasars, observed at a given SNR, taking advantage of various features observed in absorption in the spectra of bright, high-redshift quasars.

\section{The QUBRICS Survey}
Four thousand hours of observation of the ELT are a considerable investment and since 2008 the ELT shrank to 39.3 m, making the experiment even more daunting.
Besides, observations in the Southern hemisphere, where the ELT is located, risk to be hampered by the lack of luminous targets
with respect to the North, historically due to the dearth of surveys for bright quasars in the South.
This was the motivation that originated the survey QUBRICS
\citep[QUasars as BRIght beacons for Cosmology in the Southern
hemisphere]{Calderone2019,Boutsia2020}, taking advantage of the availability of several new multi-wavelength databases:
Skymapper \citep{Skymapper2019}, Gaia \citep{Gaia2021}, 2MASS \citep{2MASS2006}, WISE \citep{WISE2010}, PanSTARRS \citep{PanSTARRS2016}, DES \citep{DES2021}.
\begin{figure*}[t!]
\resizebox*{\hsize}{8.5cm}{\includegraphics[clip=true, angle=270]{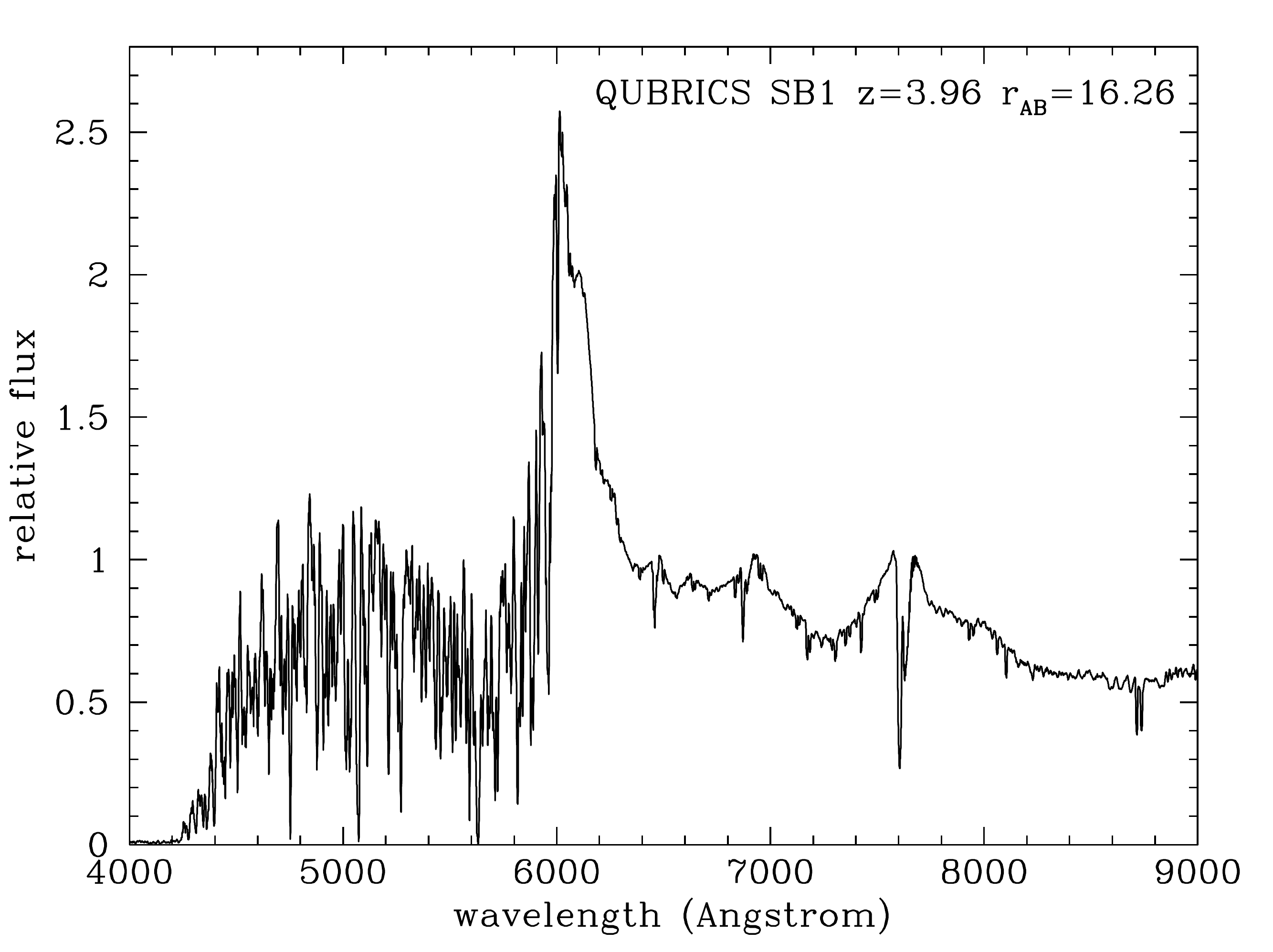}}
\caption{\footnotesize
Spectrum, obtained with MagE at the Magellan Telescope, of
a bright quasar discovered by QUBRICS, included in the Golden Sample for the Sandage test of the redshift drift (see text).
}
\label{fig:QU_SB1}
\end{figure*}
Various selection methods have been used, with particular emphasis on machine learning (ML):
in \citet{Calderone2019} candidates were selected
 using a canonical correlation analysis \citep[CCA, ][]{ref:CCA},
in \citet{Guarneri2021} the Probabilistic Random Forest \citep[PRF, ][]{Reis2019} was adopted, with modifications introduced to properly treat upper limits and missing data.
In \cite{Guarneri2022} the PRF selection was further improved, in particular adding synthetic data to the training sets. In Calderone et al., (submitted) a method, dubbed Michelangelo, has been developed to significantly boost recall
\footnote{Recall: the fraction of relevant instances (i.e., real high-$z$ QSOs) correctly classified by the algorithm. It is a statistical measure related to (but not the same as) the {\it completeness} \citep{Guarneri2022}.}
in selection algorithms, even in the presence of severely imbalanced datasets, aimed at extending the QUBRICS survey up to $z \sim 5$.

While refining the methods of selection, a continuous effort was dedicated in QUBRICS to the follow-up spectroscopy \citep{Boutsia2020}, testing the selection procedures and leading to statistically well-defined subsamples that allowed us to address the issue of the quasar luminosity function (LF) and cosmic re-ionization at $z \sim 4$ \citep{LF_Boutsia:2021ApJ...912..111B} and at $4.5<z<5$ \citep{LF_Hz_Grazian:2021arXiv211013736G}.

A strategic feature of QUBRICS is the continuous updating, after each observation cycle, of the training set, also paying attention to identify and correct the surprisingly significant fraction of erroneous spectroscopic identifications found in the literature, in order to improve the success rate and the completeness, while keeping the list of candidates manageable.

The search for high redshift Quasars is a typical "needle in a haystack" problem and is an excellent training ground for testing and developing ML techniques that are now used in a huge number of application fields. In this way it is possible to derive a series of valuable lessons such as trying to avoid the black box syndrome \citep[e.g.]{Petch2022}, heed apparently extraordinary success rates and completeness, curbe overfitting using complementary methods and dismiss stretched interpretations of non-physical features. ML is typically good for classification (i.e. giving a "label" to an object: star, galaxy, quasar...), but may be less good for regression (e.g. determining a redshift), with the interesting possibility of synergies with classical methods (e.g. model fitting) once ML has reliably identified the class of an object and therefore the model to apply.
But above all one learns that in ML training sets are the key and biases or scarcity in the training sets can produce unfair results, in facial recognition \citep{Buolamwini2018}, autonomous driving, fraud detection as well as in finding high redshift quasars. Synthetic data can be 
a useful solution in cases where real world data is limited \citep{Chaudhari2022, Guarneri2022}.

The QUBRICS survey has produced several hundred new spectroscopically confirmed bright quasars at $z>2.5$
(see Fig.~\ref{fig:z_mag_QSO}).
In \cite{Boutsia2020}
it was shown that with a new Golden Sample of 30 quasars the
redshift drift measurements, using the ANDES spectrograph at the 39m ELT, appears to be possible with less than 2500 hours of observations spread over 25 years.
New bright quasars have been found since then and new optimal observation strategies are being devised, further decreasing the required investment of time. Precursor observations with ESPRESSO have been started, that, although not detecting the drift signal, 
aim at obtaining the first statistics-limited constraint,
improving current bounds by an order of magnitude, and providing a full end-to-end proof of concept for the ANDES experiment at the ELT.
An old spectroscopist's dream is starting to come true.

\begin{acknowledgements}
SC wishes to thank Margherita Hack (and Fabio Mardirossian) for having nurtured in Trieste an environment marked by freedom and originality, without which it would have not been possible for him to arrive in this city and develop in a gratyfing way his humble research.
\end{acknowledgements}


\bibliographystyle{aa}
\bibliography{00Cristiani_Hack100} 

\end{document}